%% file: main.tex
\documentclass[a4paper,11pt]{article}
\usepackage{aaskaiid}
\usepackage{orcidlink}

\graphicspath{{figures/}}

\title{The SKA as a Probe of Heliospheric Turbulence and Radio-wave Propagation Effects}
\ShortTitle{Probing heliospheric turbulence}

\author[1,2]{Nicolina Chrysaphi \orcidlink{0000-0002-4389-5540}}
\ShortName{Chrysaphi et al.} 
\author[2]{Eduard P. Kontar \orcidlink{0000-0002-8078-0902}}
\author[3]{Hamish A. S. Reid \orcidlink{0000-0002-6287-3494}}
\author[4]{Sophie Musset \orcidlink{0000-0002-0945-8996}}
\author[5]{Shilpi Bhunia}
\author[6]{Kamen Kozarev}

\affiliation[1]{INAF -- IAPS, Via del Fosso del Cavaliere, 100, 00133 Rome, Italy}
\emailAdd{nicolina.chrysaphi@inaf.it}
\affiliation[2]{School of Physics \& Astronomy, University of Glasgow, Glasgow, G12 8QQ, UK}
\affiliation[3]{Mullard Space Science Laboratory, University College London, Holmbury St. Mary, Dorking, Surrey, RH5 6NT, UK}
\affiliation[4]{Johns Hopkins Applied Physics Laboratory, 11100 Johns Hopkins Road, 20723, Laurel, MD, USA}
\affiliation[5]{LIRA, Observatoire de Paris, Université PSL, CNRS, Sorbonne Université, Université de Paris, 5 place Jules Janssen, 92195 Meudon, France}
\affiliation[6]{Institute of Astronomy and National Astronomical Observatory - Bulgarian Academy of Sciences, 1784 Sofia, Bulgaria}

\abstract{
Density turbulence in the heliosphere can impact traversing radio photons originating from anywhere in the universe, leading to distortions of both the spectroscopic and imaging properties of the radio sources. Extra-solar radio sources exhibit scintillation and angular broadening, while monochromatic signals from spacecraft are spectrally broadened. It was recently demonstrated that such impacts, referred to as radio-wave propagation effects, are particularly significant in observations of solar radio bursts excited through the plasma emission mechanism. A comparison of detailed observations with simulations is required in order to quantify the radio-wave propagation effects and disentangle the true radio-source properties from the observed ones, enabling the diagnosis of the heliospheric environment. Consequently, theoretical advancements and our ability to quantify the heliospheric turbulence depend on the quality and quantity of available observations. SKA pathfinders, like LOFAR, have been key to the significant progress recently achieved, but have also highlighted areas where the available observing capabilities are lacking. The SKA's unrivalled sensitivity will be crucial in identifying and distinguishing fine and faint radio structures, where our ability to model them defines whether we can accurately describe the heliospheric turbulence and deduce the fundamental properties of the radio sources. The SKA will be an indispensable tool in observing solar radio emissions from the Sun to beyond 1 au, complemented by ground-based interferometers that can reach frequencies down to the ionospheric cut-off at 10 MHz and space-based radio instruments which cover frequencies down to a few kHz.
}

\begin{document}
\maketitle

\include{journal-names}

\section{Introduction} \label{sec:intro}
Solar matter constantly escapes the Sun's gravitational pull and flows outwards (giving rise to the solar wind) until it encounters the interstellar medium, forming a termination shock. This shock defines the edge of the \emph{heliosphere}, the area under the control and influence of the Sun. The solar wind is turbulent, generating density fluctuations of various scales throughout the heliosphere. Given that the heliospheric environment is populated by density inhomogeneities, any photons that traverse the heliosphere will encounter these inhomogeneities as they propagate towards the detector, regardless of whether it is ground- or space-based. Most types of radiation do not interact strongly with the heliospheric density turbulence, and thus reach our detectors unaffected by the heliospheric environment through which they have propagated. This is not the case for radio photons, the propagation of which can be heavily influenced by such density inhomogeneities through processes collectively known as \emph{radio-wave propagation effects}, which include scattering (i.e., random, small-angle trajectory deviations).

The interaction between radio waves and density inhomogeneities has long been studied through the alterations that it induces on the observed properties of various radio sources (presented in Section~\ref{sec:diagnosing_heliosphere}). Such radio sources include extra-solar radio emitters, radio signals emitted by spacecraft, and solar radio bursts. Although they have different (astrophysical) origins and result from different emission mechanisms, all these radio waves interact with---and thus respond to---the very same heliospheric turbulence, meaning that their observed properties reflect the heliospheric environment, enabling its remote diagnosis \citep[e.g.,][]{2023ApJ...956..112K}. These radio signals traversing the heliosphere are a diagnostic of heliospheric turbulence at ion scales, since the radio-wave scattering rate depends on the ion-scale density turbulence \citep[e.g.,][]{2025ApJ...991L..57K}.

Recent advances in our ability to perform both spectroscopic and imaging observations of radio sources have enabled substantial improvements in the modelling and quantification of these propagation effects (Section~\ref{sec:diagnosing_heliosphere}). In this chapter, we give an overview of how the upcoming Square Kilometre Array (SKA) observations will enable further advances in the quantification and understanding of these propagation effects, and thus the heliospheric environment itself. The major improvement is anticipated to occur due to the considerably higher sensitivity of SKA measurements (Section~\ref{sec:SKA_advantages}). We also highlight the critical role that the SKA will play as part of a synergy of radio observatories, with a particular focus on the coordination with space-based instruments (Section~\ref{sec:synergies}).

\section{Diagnosing the heliospheric environment using radio observations} \label{sec:diagnosing_heliosphere}
This section aims to give a brief overview of the diagnostic power of radio observations. It should not be viewed as a review, and the included references are by no means exhaustive, nor do they necessarily reflect the state of the art of developed methods. Nevertheless, we introduce a variety of radio measurements that can be used to remotely probe the heliospheric environment, alluding to the wide range of studies that can be conducted with the SKA.

\subsection{Extra-solar radio sources} \label{sec:extra-solar_sources}
\begin{figure}[ht]
    \centering
	\includegraphics[width=0.6\columnwidth]{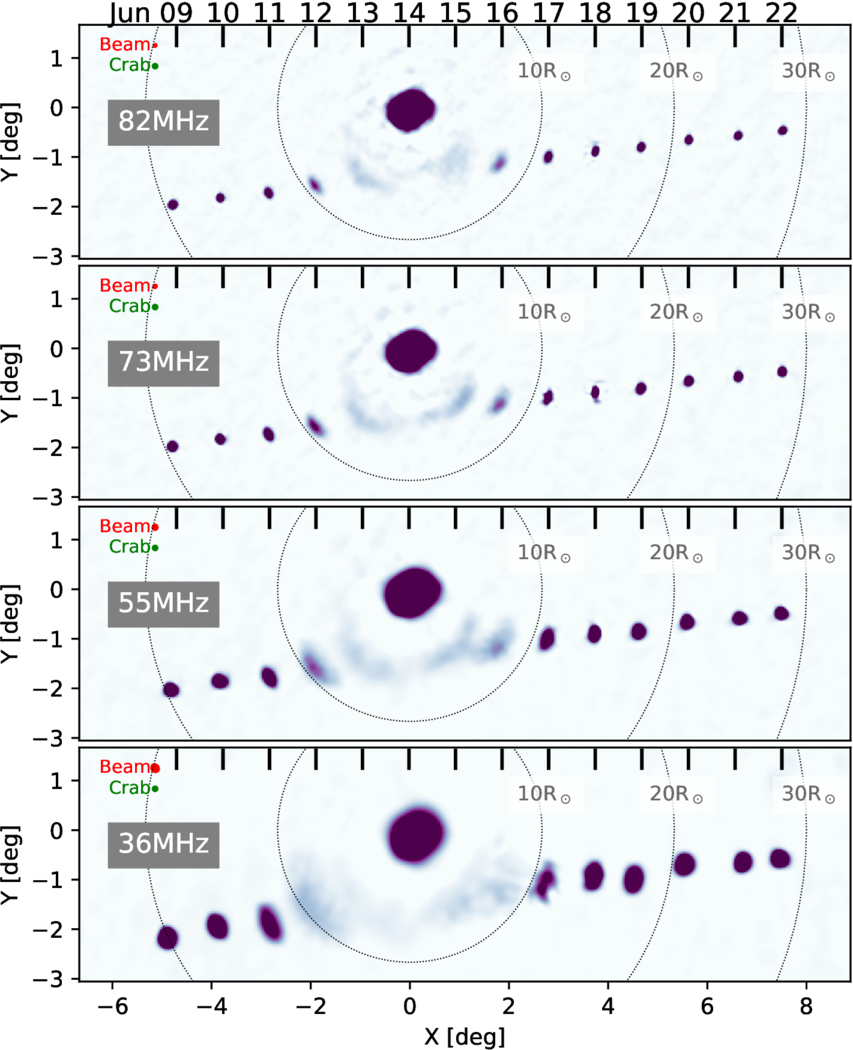}
    \caption{
    Images of the Crab Nebula (Tau A) taken by the Owens Valley Radio Observatory Long Wavelength Array (OVRO-LWA) from 9--22 June 2024 (indicated by the top x-axis) at four different frequencies (36, 55, 73, and 82 MHz) during its solar transit. Daily images were taken at a fixed time (near 12:00 at the location of the interferometer). The image of the Sun (centred at (0,0) in each panel) was taken on 12 June 2024. The red and green circles at the top right of each panel delineate the full-width at half-maximum (FWHM) size of the synthesised beam and the true size of the Crab Nebula, respectively, at each frequency. It is evident that for every frequency measurement, the apparent size of the Crab Nebula is significantly larger than both the beam and its true size, while its size broadens and becomes more elliptical when it is closer to the Sun. For each given position in its transit, the Crab Nebula also appears larger at decreasing frequencies. All panels show the attenuation of the Crab Nebula by the solar corona once it reaches a heliocentric distance of $\lesssim$10 solar radii ($\mathrm{R_\odot}$), with the onset of the attenuation seen at larger heliocentric distances for lower frequencies.
    Figure taken from \cite{2025ApJ...992..128Z} and reproduced under \href{https://creativecommons.org/licenses/by/4.0/}{CC BY 4.0}.
    }
    \label{fig:crab_nebula}
\end{figure}

The interference of radio waves by the heliospheric environment was appreciated relatively early in the 20th century, and was confirmed through observations of compact extra-solar radio emitters \citep[effectively point-sources;][]{1952Natur.170..319M}. In this work it was observed that, as a compact extra-solar source (in this case the Crab Nebula) approaches the Sun in the plane of sky, the solar corona attenuates (or occults) its radio emissions, leading to gradual decreases in intensity as a function of heliocentric distance, and eventually entirely prevents its observation. The observed source size was also found to broaden as the source's angular separation from the Sun decreased (see Fig.~\ref{fig:crab_nebula}). Such variations in the observed intensity of compact radio sources as a function of heliocentric distance have been attributed to scattering off small-scale density inhomogeneities in the solar corona \citep{1955RSPSA.228..238H}, allowing for approximations of the size and density of these inhomogeneities. \cite{1958MNRAS.118..534H} further studied the angular broadening of an extra-solar point source as a function of its angular separation from the Sun, confirming that the source broadened with decreasing separation, and also demonstrating the ellipticity of the sources. This led to the (now-established) suggestion that anisotropic scattering must be at play. Specifically, it was argued that density inhomogeneities are aligned with and elongated along the magnetic field, leading to scattering that is stronger in the perpendicular (to the magnetic field) direction compared to the parallel one. This then enabled the diagnosis of the magnetic field direction at the heliocentric distances where the elliptical extra-solar sources were recorded \citep{1958MNRAS.118..534H}.

Scintillation (i.e., rapid fluctuations in the recorded signal amplitude) of various radio sources---independent of the intrinsic emission of the radio sources---was also observed \citep{1964Natur.203.1214H}. Although such amplitude scintillations can result from perturbations affecting the refractive index in the Earth's ionosphere, many of the observed scintillations were deemed to be caused by density inhomogeneities in the heliosphere. Extra-solar sources have been used to demonstrate the ways in which the heliospheric radio-signal interference differs from that of the ionosphere.
For example, the scintillation of the radio sources depends on the angular separation between the source and the Sun, in line with the impact of propagation through the denser parts of the solar atmosphere (which also causes the angular broadening and attenuation of the radio emissions), while the time scale of the scintillations is significantly shorter (in the order of $\sim$1~s or even sub-seconds) than what ionospheric disturbances can account for \citep{1964Natur.203.1214H}. In addition to being one of the ways to investigate the properties of heliospheric density inhomogeneities, scintillation measurements alone (i.e., in the absence of (adequate) imaging observations) can be used to estimate the angular diameter of small radio sources \citep{1964Natur.203.1214H}.

Scintillation of the signal amplitude induced by radio-wave propagation through heliospheric density inhomogeneities has also been observed for various other radio sources (beyond compact extra-solar sources), including Jovian radio emissions \citep{1967ApJ...148..885D}.

\subsection{Synthetic radio sources} \label{sec:synthetic_sources}
Heliospheric density fluctuations are also known to affect narrow-band (monochromatic) signals transmitted by spacecraft, occulting them when the spacecraft move behind the Sun in their orbit (as observed from the plane of the receiving antenna, and analogously to the attenuation of natural radio sources such as the Crab Nebula), while also causing them to spectrally broaden \citep[see Fig.~\ref{fig:all_radio_obs}(e);][]{1969Sci...166..598G, 2024ApJ...968...72A}. Moreover, a phase-difference scintillation (not to be confused with amplitude scintillation) is observed when signals of different frequencies are transmitted from the same spacecraft \citep{1976ApJ...210..568W}. Such spacecraft radio transmissions enable the study of a large range of scale sizes of heliospheric density fluctuations, as well as a large part of the heliosphere itself. \cite{1979JGR....84.7288W} used phase scintillation and spectral broadening of high-frequency signals (a few GHz) transmitted by spacecraft to ground-based receivers to probe distances from $\sim$2--215~$R_\odot$ (where the distance between the Sun and Earth is $\sim$215~$R_\odot =1~\mathrm{au}$).

\subsection{Solar radio emissions} \label{sec:solar_emissions}
Like many other astrophysical objects, the Sun is a constant emitter of radio radiation, but also the source of violent, sporadic eruptions exciting radio waves that are much more intense than any other radio signal traversing the heliosphere, including the radio radiation emitted constantly by the relatively-quiet Sun. The most prominent solar activities producing intense radio emissions are solar flares and shocks driven by Coronal Mass Ejections \citep[CMEs; e.g.,][]{2014RAA....14..773R, 2020ApJ...893..115C}. They can accelerate semi-relativistic electrons that propagate through the heliosphere and excite radio waves---referred to as \emph{solar radio bursts}---via the coherent plasma emission mechanism \citep{1987SoPh..111...89M}. The emission frequency of such radio bursts is very close to the local electron plasma frequency of the heliosphere, which depends on the local electron density. This intrinsic relation makes solar radio bursts particularly susceptible to heliospheric density fluctuations and, by extension, a powerful diagnostic of the heliospheric environment. Solar radio bursts often extend over a relatively large range of frequencies, but also display various fine structures (in frequency and time) that result from the presence of heliospheric density fluctuations \citep{2021NatAs...5..796R, Kontar01.2026.SKA}, requiring high-resolution measurements to resolve, which the SKA will provide (Section~\ref{sec:imaging_spectroscopy}). Moreover, solar radio bursts are ubiquitous events; they tend to occur numerous times a day, across various longitudes, and across various frequencies, maximising their diagnostic potential. Although the regime under which solar radio photons are scattered differs from that of other radio sources \citep[such as extra-solar sources; e.g.,][]{2023ApJ...956..112K}, they all respond to the same density inhomogeneities and thus exhibit changes that reflect the properties of the very same heliospheric environment. In the case of solar radio bursts, understanding and quantifying these changes has been somewhat challenging, constituting a topic of debate for several decades.

Recent state-of-the-art observations have re-ignited the interest in investigating and understanding the possible impacts of propagation effects on solar radio emissions. In particular, imaging spectroscopy observations by the LOw-Frequency ARray \citep[LOFAR, an SKA pathfinder;][]{2013A&A...556A...2V} encouraged the examination of solar radio bursts and their fine structures in unprecedented detail, revealing the sub-second evolution of their properties \citep[e.g.,][]{2017NatCo...8.1515K, Kontar01.2026.SKA}. LOFAR observations were used to quantitatively demonstrate the significance of scattering effects on observed radio properties and, crucially, their physical interpretations \citep{2017NatCo...8.1515K, 2018ApJ...868...79C}. Radio-wave propagation effects (predominantly scattering) have been quantitatively shown to impact the majority of solar radio burst observables, masking their true properties: source sizes are broadened and their ellipticity---and thus measured size---depends on the observer's position \citep{2019ApJ...884..122K, 2020ApJ...898...94K}; sources are displaced from their true location \citep{2018ApJ...868...79C, 2023A&A...680A...1C, 2025A&A...696A.124C}; the measured flux varies with the observer's position \citep{2021A&A...656A..34M}; both the decay and rise phases of the intensity-time profiles broaden \citep{2019ApJ...884..122K, 2024A&A...687L..12C}; and the frequency ratio between fundamental and harmonic components is lower than theoretical predictions \citep{2023MNRAS.520.3117C}.

The necessity to take into account anisotropic scattering off small-scale density fluctuations to \emph{simultaneously} reproduce multiple observed properties of solar radio bursts was first illustrated by \cite{2019ApJ...884..122K}, using the state-of-the-art radio-wave propagation simulations they developed. Several studies followed, in which multiple observed radio burst properties (and even their sub-second evolution) were reproduced simultaneously. These studies addressed various persistent open questions and quantitatively confirmed that anisotropic scattering off small-scale density inhomogeneities dominates the measured properties and thus cannot be neglected \citep{2020ApJ...898...94K, 2020ApJ...905...43C, 2023MNRAS.520.3117C}. Key to the success of these studies was LOFAR's ability to resolve fine-size, faint radio structures, both spectroscopically and in imaging observations. The successful, simultaneous reproduction of multiple solar radio burst properties by radio-wave propagation simulations is a vital achievement, as it enables the diagnosis of the level of turbulence and anisotropy in the heliospheric environment itself \citep{2020ApJ...905...43C, 2020ApJ...898...94K, 2021A&A...656A..34M, 2023ApJ...956..112K, 2023MNRAS.520.3117C, 2025ApJ...978...73C}. The accurate characterisation of heliospheric turbulence is important not only for the ability to correct the substantial distortions induced in radio observations traversing the heliosphere, but also for the unique exploitation of radio observations for inferring fundamental properties of the heliospheric environment and the violent solar eruptions that excite these emissions. For example, it was recently discovered that, due to scattering off anisotropic density fluctuations, solar radio burst photons emitted from a given location (i.e., a single radio source) will trace the magnetic field as they propagate out into the heliosphere, revealing the configuration of the interplanetary magnetic field \citep{2025NatSR..1511335C}.
By combining radio observations of solar radio bursts with in-situ spacecraft measurements, \cite{2025ApJ...991L..57K} inferred the properties of density and magnetic-field fluctuations across a wide range of heliocentric distances, from $\sim$10~$\mathrm{R_\odot}$ to $\sim$1~$\mathrm{au}$ (where the in-situ measurements are available). They demonstrate that Kinetic Alfv\'en Waves (KAWs) are consistent with observed magnetic and density fluctuations, suggesting that KAWs play a central role in the energy cascade and the radio-wave scattering in the heliosphere.

\begin{figure}[ht]
    \centering
	\includegraphics[width=1.0\columnwidth]{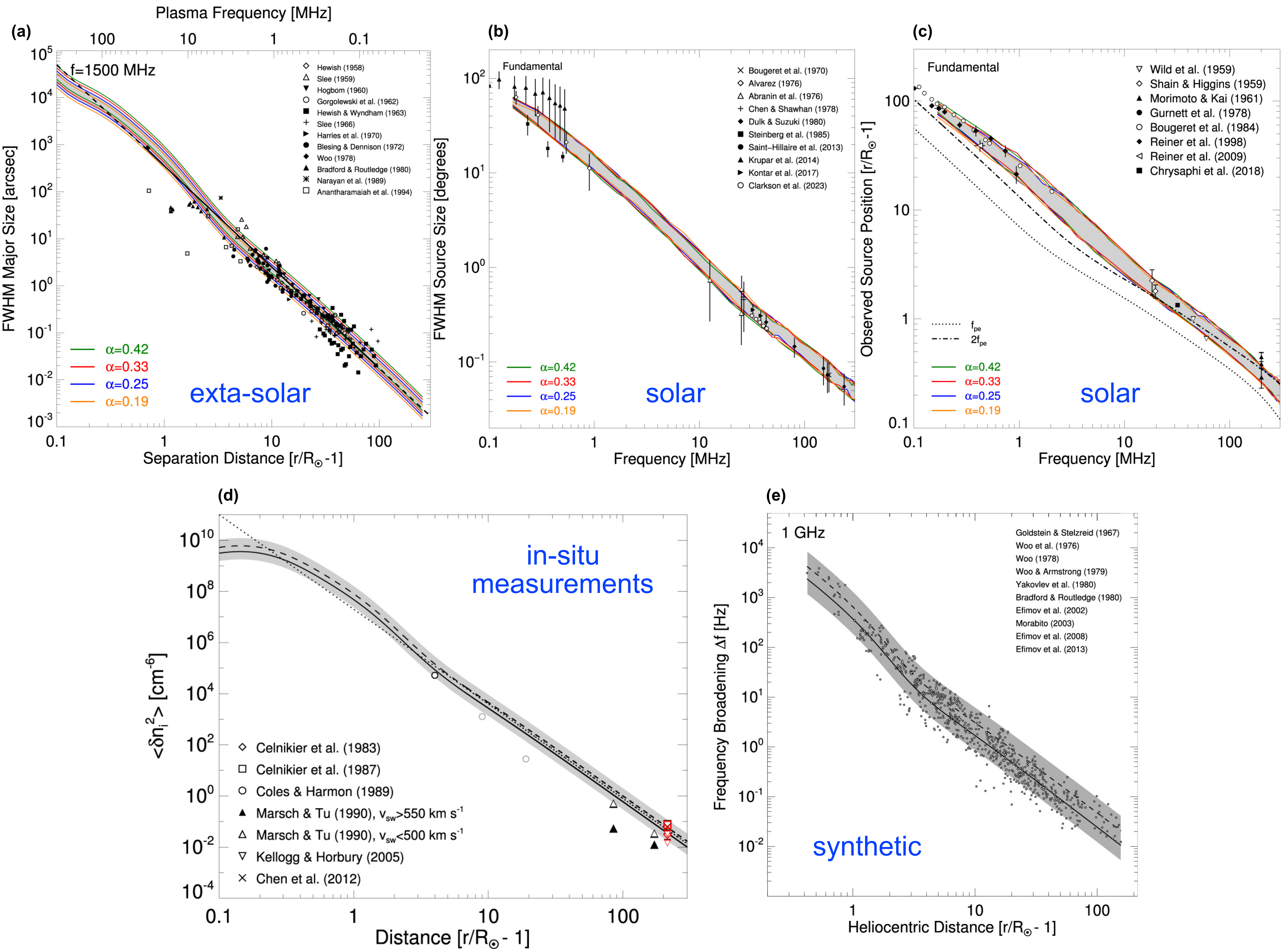}
    \caption{
    A multitude of observations---of (a) extra-solar radio sources, (b)--(c) solar radio sources, (d) in-situ density measurements, and (e) synthetic radio sources---that have all been successfully reproduced using the same, narrow range of anisotropy values describing the heliospheric density inhomogeneities. The measurements cover a large range of heliospheric distances, from the Sun to 1~$\mathrm{au}$.
    The observations consistently indicate that the average heliospheric level of density turbulence is $\sim$2.5--4 times stronger in the perpendicular-to-the-magnetic-field direction compared to the parallel one.
    Figures were taken and adapted from \cite{2023ApJ...956..112K} (panels~(a)--(d)) and \cite{2024ApJ...968...72A} (panel~(e)) under \href{https://creativecommons.org/licenses/by/4.0/}{CC BY 4.0}.
    }
    \label{fig:all_radio_obs}
\end{figure}

\cite{2023ApJ...956..112K} used a single anisotropic propagation model to successfully describe observations of solar radio bursts, extra-solar sources, and in-situ spacecraft measurements of heliospheric turbulence. By combining observations over several decades, a range of values was inferred that describe the \emph{average} characteristics of heliospheric turbulence from the Sun to 1~$\mathrm{au}$ (see Fig.~\ref{fig:all_radio_obs}). The challenge lies in unlocking the ability to accurately describe the heliospheric environment on a regular basis, including its reaction to various transient solar events and its evolution with distance and time. Radio emissions (of solar, extra-solar, or synthetic origins) are uniquely positioned to do that (Fig.~\ref{fig:all_radio_obs}), providing a window into the ways that solar---and by extension stellar---activity affects the surrounding environment. This enables constraining local environmental properties \citep[e.g.,][]{2023ApJ...956..112K, 2024MNRAS.527.9872G, 2024ApJ...968...72A}, and improves understanding of the mechanisms accelerating energetic particles in astrophysical environments. A key requirement for reaching this milestone is the availability of highly-resolved spectroscopic and imaging observations that will facilitate the complete coverage of the heliospheric environment, with the SKA being a crucial piece of this puzzle (see Sections~\ref{sec:SKA_advantages} and \ref{sec:synergies}).

\section{The advantages of SKA observations} \label{sec:SKA_advantages}
In Section~\ref{sec:diagnosing_heliosphere}, we presented a variety of radio observations that can be conducted to remotely probe the heliospheric turbulence. We also indicated how the observing capabilities of LOFAR, a pathfinder of the SKA, enabled a robust quantitative comparison of solar radio burst observations with models of radio-wave propagation through the heliosphere.
The SKA's improved observing capabilities will facilitate more advanced examinations of radio-wave propagation effects, and thereby a more accurate description of the dynamic heliospheric environment.
In this section, we describe the SKA's four major observational strengths that will be particularly useful in studies of the heliospheric turbulence: its sensitivity, its ability to perform high-resolution imaging spectroscopy, its frequency range, and its geographic location.

The SKA consists of two distinct telescopes, each covering a different frequency range. These are the SKA-Low (50--350~MHz), comprised of dipole antennas and located in Australia, and the SKA-Mid (0.35--15.4~GHz), comprised of dishes and located in South Africa.
The currently-defined vision of the SKA (also known as ``SKA phase 1'', or SKA1) will be deployed in five stages, known as Array Assemblies (AA), with the last two corresponding to the operational phase. The final SKA1 stage is called \emph{Design Baseline (AA4)}, but since funding is yet to be entirely secured for AA4, a construction pause is planned at the \emph{Staged Delivery (AA*)}, which is currently expected to be commissioned by 2029, marking the beginning of the operational phase. The specifications for AA4 and AA* can differ, providing different observing capabilities, as will be outlined in the following sections.

\subsection{Sensitivity} \label{sec:sensitivity}
The major enhancement that the SKA offers to our current observing abilities is its sensitivity, surpassing that of any other existing instrument by a considerable amount. The increased sensitivity will allow previously-unresolved faint and fine radio structures to be observed, both spectroscopically and in imaging observations, benefiting all radio measurements, whether the origin of the signal is solar, extra-solar, or synthetic. A higher sensitivity (when combined with sufficient temporal, spectral, and spatial resolutions) leads to a more accurate characterisation of sources (size and location) as well as signal amplitude (including scintillations and attenuation), which constitute crucial information for the diagnosis of heliospheric turbulence. Moreover, the improved sensitivity will facilitate a more accurate discrimination and identification of various signals, whether several solar radio bursts occur at the same time and frequency, or whether observations of extra-solar or synthetic sources are polluted by solar emissions or other radio interference.

The number of SKA-Mid dishes will increase from 144 at AA* to 197 at AA4, expanding the maximum SKA-Mid baseline from $\sim$108~km to $\sim$159.6~km, increasing both its angular resolution and sensitivity. The maximum baseline of SKA-Low remains effectively the same at AA* and AA4 ($\sim$73.4~km). However, the number of SKA-Low stations will increase from 307 at AA* to 512 at AA4, improving the $uv$-coverage and sensitivity of measurements. For comparison, the sensitivity of SKA-Low at AA4 will be approximately 8 times higher than that of LOFAR at equivalent frequencies (50--90, 110--240~MHz), with the factor depending on the frequency and observing parameters.

\subsection{High-resolution imaging spectroscopy} \label{sec:imaging_spectroscopy}
The increased sensitivity that the SKA offers would be inconsequential if it were not accompanied by very high temporal, spectral, and spatial resolutions. The ability to spectroscopically resolve the very faint and fine structures is of outmost importance when it comes to using radio emissions to diagnose heliospheric turbulence (Section~\ref{sec:diagnosing_heliosphere}). For example, a statistical analysis of the fine structures (striae) of Type IIIb solar radio bursts at frequencies ($f$) from 30--80~MHz showed that they have an average relative instantaneous bandwidth $\Delta f/f \approx 0.1$\% and durations as short as $\sim$0.5~s \citep{2018SoPh..293..115S}. Additionally, radio-wave propagation effects are responsible for very short time-scale variations of spectroscopic and imaging radio properties (in the orders of $\sim$1~s or sub-seconds), for extra-solar, synthetic, and solar radio sources (Section~\ref{sec:diagnosing_heliosphere}). Moreover, the apparent duration of solar radio bursts is frequency dependent. This is particularly evident in Type III bursts, whose intensity-time profiles become increasingly narrower with increasing frequency \citep[e.g.,][]{2014RAA....14..773R, 2024A&A...687L..12C}, requiring higher temporal resolutions to resolve. 

High-resolution imaging spectroscopy observations with the SKA will be extremely advantageous when compared to simulations of imaging and spectroscopic properties at sub-second scales.  This has already been demonstrated by various studies of solar radio busts using LOFAR observations \citep[Section~\ref{sec:solar_emissions};][]{2020ApJ...898...94K, 2020ApJ...905...43C}.  Imaging spectroscopy enables imaging of the radio emissions at every frequency and time for which the spectrum is recorded, meaning that for every pixel on a dynamic spectrum, there exists an image of the radio source(s), facilitating the examination of the evolution of emissions as a function of time or frequency.

As outlined in Section~\ref{sec:solar_emissions}, LOFAR observations have been instrumental in the recognition that solar radio bursts are distorted by heliospheric density fluctuations and can thus be exploited as a diagnostic tool of heliospheric turbulence. LOFAR's tied-array observing mode provided simultaneous spectroscopic and imaging observations with equivalent, high resolutions that resolved fine solar radio burst structures. Several studies that focused on quantifying radio-wave propagation effects, utilised tied-array observations by LOFAR's outer-core Low-Band-Antenna stations ($\sim$10--90~MHz) with a maximum baseline of $\sim$3.5~km, recording emissions with a temporal resolution of 0.01~s and a spectral resolution of 12.2~kHz \citep[e.g.,][]{2018ApJ...868...79C, 2020ApJ...905...43C, 2020ApJ...898...94K}.

The SKA will also deliver sub-second temporal resolution, the exact value of which is defined by the observing mode and parameters.
A noticeable improvement in observations will also arise from the SKA's higher spectral resolution. In the (interferometric) ``Continuum'' mode, SKA-Low will be able to record at 5.43~kHz intervals, while SKA-Mid will record every 13.4~kHz. The integration times (i.e., temporal resolutions) for this mode are 0.85~s and 0.125~s for SKA-Low and SKA-Mid, respectively. The integration time of SKA-Low can be reduced to 0.3~s, but for a limited number of sub-stations.
Much higher spectral resolutions can be reached for a certain portion of the recorded bandwidth when the SKA is set to operate in the ``Zoom/Spectral line'' mode. In Zoom mode, SKA-Low will be able to record every 
0.0141~kHz for a bandwidth of 24.4~kHz and up to every 1.808~kHz for a bandwidth of 3125.2~kHz, whereas SKA-Mid will be able to record every 0.21~kHz for a bandwidth of 3.125~MHz and up to every 13.44~kHz for a bandwidth of 200~MHz. Similar to LOFAR, the SKA has beamforming capabilities, allowing both of its telescopes to form tied-array beams, providing impressive temporal resolutions. When in the beamforming ``Pulsar Search'' mode, SKA-Low will be able to record every 69~$\mathrm{\mu s}$ and 14.4~kHz with a maximum bandwidth of 118.5~MHz, and SKA-Mid will be able to record every 65~$\mathrm{\mu s}$ and 107.5~kHz with a maximum bandwidth of 300~MHz. The maximum baseline for both SKA telescopes when set to any beamforming Pulsar modes is 20~km (at both AA* and AA4). The maximum number of beams formed during the Pulsar Search mode at AA* is 250 for SKA-Low and 1125 for SKA-Mid, and at AA4 it will increase to 500 and 1500 beams, respectively.
With such high temporal and spectral resolutions, together with the increased sensitivity, the SKA will produce spectra that depict the fine structures of solar radio bursts in unprecedented detail, contributing to more accurate descriptions of the radio-source properties and the quantification of radio-wave propagation effects.

Although the maximum baseline of the SKA is inferior to the maximum baseline of LOFAR, the SKA will nevertheless be able to resolve solar radio burst sources at corresponding frequencies. Different subarray templates have been defined for the SKA, allowing for observations with various sets of stations and baselines. To provide spatial resolutions that surpass those of LOFAR's outer-core typically utilised in radio-wave propagation investigations of solar emissions (3.5~km baseline), SKA-Low observations at AA4 will need to be conducted---as a minimum---using the subarray that contains all stations within a radius of 3~km from the array centre, corresponding to a 5~km baseline. To achieve higher spatial resolutions than LOFAR at the AA* stage as well, the SKA-Low subarray containing stations within 6~km from the centre will need to be used (as a minimum), delivering an 8.5~km baseline (at both AA* and AA4). For comparison, at 70~MHz, the nominal spatial resolution of LOFAR with a baseline of 3.5~km is $\sim$4.2~arcmin, whereas with a baseline of 8.5~km the SKA will provide a spatial resolution of $\sim$1.73~arcmin, corresponding to an improvement factor of $\sim$2.4.
Using the entire 20~km baseline available for beamformed Pulsar-Search observations will provide the optimal observing conditions in terms of the temporal, spectral, and spatial resolution ($\sim$0.74~arcmin at 70~MHz, a factor of $\sim$5.7 higher than LOFAR observations with a 3.5~km baseline).

\subsection{Frequency range} \label{sec:freq_range}
The SKA will record radio emissions over a very large frequency range. Specifically, SKA-Low will cover frequencies from 50--350~MHz, whereas SKA-Mid will collectively cover frequencies from 350~MHz to 15.4~GHz. SKA-Mid is divided into six bands, four of which are part of the AA* stage (covering frequencies from 0.35--1.76~GHz and from 4.6--15.4~GHz), and the other two bands are part of the AA4 stage (covering the intermediate frequencies of 1.65--5.18~GHz).

The SKA frequency range at AA* will already suffice for conducting measurements of all radio signals relevant to the diagnosis of heliospheric turbulence, whether they originate from solar, extra-solar, or synthetic sources (Section~\ref{sec:diagnosing_heliosphere}). Synthetic signals from spacecraft are transmitted at frequencies $\gtrsim$1~GHz---well-above the heliospheric plasma frequency level---meaning that they will be measured by SKA-Mid (0.35--15.4~GHz). However, compact extra-solar radio sources tend to be observed at much lower frequencies, within the SKA-Low range (50--350~MHz), although measurements in the GHz range have also been conducted \citep[e.g.,][]{1994JApA...15..387A}. In terms of solar observations, the higher frequencies recorded by SKA-Mid will enable the localisation and tracking of the solar features from which the accelerated particles exciting the radio emissions originate. On the other hand, SKA-Low will be particularly useful for the purposes of observing solar radio emissions resulting from the plasma emission mechanism, which are the ones affected by---and thus useful for probing---heliospheric turbulence, since these emissions tend to be observed below $\sim$1~GHz, and predominantly below $\sim$500~MHz. However, solar radio bursts can have a relatively large bandwidth of continuous emissions, for example, from $\sim$500~MHz down to $\sim$200~MHz. Hence, for solar studies, synchronous observations with the SKA-Low and SKA-Mid telescopes would be ideal, although practically difficult due to their vastly different locations (Section~\ref{sec:location}). For this purpose, synergies with other instruments would be prudent (Section~\ref{sec:synergies}).

A narrow observational gap exists in the lower frequencies (from $\sim$10--50~MHz), since SKA measurements do not extend down to the ionospheric cut-off at $\sim$10~MHz. However, this gap can be filled by coordinating with other available radio instruments, both ground- and space-based (see Section~\ref{sec:synergies}). It is worth noting that when it comes to ground-based instruments, these lower frequencies tend to be highly susceptible to radio-frequency interference, frequently making them impractical for scientific use, especially for instruments of high sensitivity like the SKA. As such, for the purposes of fully tracking solar radio bursts and probing heliospheric turbulence, it will be valuable to complement SKA frequencies with those recorded by space-based detectors.

\subsection{Geographic location} \label{sec:location}
The SKA-Low telescope is located in Australia, whereas SKA-Mid is located in South Africa. Thereby, the SKA Observatory (SKAO) is located in the Southern Hemisphere, capturing a different part of the sky than the telescopes based in the Northern Hemisphere (like LOFAR). Whilst this means that synchronous observations of the same events with telescopes in the Northern Hemisphere may not be possible, it also functions as a strength when it comes to complementing SKA observations with other instruments (see Section~\ref{sec:synergies}). Considering heliospheric studies, the ability of a telescope to resolve the Sun or sources in the heliosphere varies with the time of year, fluctuating between the Northern and Southern Hemispheres, thanks to the varying elevation of the Sun in the particular location (which is always optimal during the corresponding summer months in each hemisphere). For example, in December (and the months around it), when heliospheric observations in the Northern Hemisphere are substandard, the conditions in the Southern Hemisphere are at their most favourable. The SKA telscopes will also be located closer to the equator than many other radio interferometers (including LOFAR), which is an additional advantage for heliospheric studies, given that the ability to resolve solar radio sources depends on the elevation of the Sun in the sky \citep{2022ApJ...925..140G}, which is maximised for observations at noon and the closer a telescope is to the equator. Moreover, the chosen sites for the SKA telescopes are relatively radio-quiet compared to many other telescope sites around the world. This is important for the quality of the observations, especially when measurements of high sensitivity are conducted (Section~\ref{sec:sensitivity}), but also in terms of minimising radio-frequency interference polluting astronomical observations, something crucial in an era where increasingly more noise from artificial sources is detected \citep[including from satellites;][]{2025A&A...698A.244Z}.

\section{Synergies with other instruments} \label{sec:synergies}
Synergies between the SKA and other radio instruments are beneficial to the heliophysics community for two main reasons: (i) the location of the SKA in the Southern Hemisphere, and (ii) the large frequency range it covers. Coordination with telescopes based in the Northern Hemisphere means that the Sun and its heliosphere can be observed under optimal conditions for longer periods during a year (as described in Section~\ref{sec:location}), providing year-long observations that resolve the various emissions and enable a more continuous and accurate tracking of their evolution. The amount of daylight that each hemisphere receives also varies with seasons, so telescopes in the two hemispheres can complement each other. Depending on the conditions, the SKA could also observe simultaneously with telescopes in the Northern Hemisphere. For example, SKA-Mid could observe at the same times as LOFAR (which covers 10--90, 110--240~MHz), complementing each other's frequencies. Some other instruments in the Northern Hemisphere which could work in synergy with the SKA include OVRO-LWA ($\sim$13--87~MHz), the Karl G.\ Jansky Very Large Array \citep[VLA; 0.054--50~GHz;][]{2011ApJ...739L...1P}, and the solar-dedicated Mingantu Spectral Radioheliograph \citep[MUSER; 0.4--15~GHz;][]{2023...Yan...MUSER} which will be expanded to cover 30--400~MHz as well.

Although there exist a plethora of radio instruments around the world, instruments with advanced observing capabilities are required to enable the diagnosis of the heliospheric turbulence. Instruments complementing SKA observations should deliver spectra with high temporal and spectral resolutions, and ideally be accompanied by resolved images. In the case of ground-based detectors, sub-second resolutions are necessary to depict the variations in the radio properties caused by heliospheric turbulence (Section~\ref{sec:diagnosing_heliosphere}). For the case of solar radio emissions, high-cadence (sub-second) images are also required for a robust analysis.

Due to its extensive frequency range (0.05--15.4~GHz; Section~\ref{sec:freq_range}), the SKA alone will be able to observe most of the radio emissions of interest that ground-based instruments can record (whether solar, extra-solar, or synthetic). Earth's ionosphere prevents any radio signals below $\sim$10~MHz from reaching the ground, defining the ionospheric cut-off. Given that the lowest frequency the SKA will observe is 50~MHz, synchronous observations with other telescopes (primarily in the Southern Hemisphere, if comparable imaging conditions are to be maintained) that can record down to $\sim$10~MHz may be desired depending on the observational objectives. Space-based instruments, on the other hand, are not restricted by the ionosphere and can observe much lower frequencies. Solar-dedicated space-based radio detectors tend to observe frequencies from $\sim$20~MHz down to a few kHz, meaning that they can complement the frequency range of the SKA. Space-based detectors are a crucial piece of the puzzle for heliospheric studies since solar radio bursts are regularly observed below $\sim$10~MHz, and spacecraft can (in principle) take nearly-continuous measurements as they are not limited by night-time observations. Prominent examples of current solar-dedicated spacecraft with radio instruments include Solar Orbiter \citep[][]{2020A&A...642A...1M, 2020A&A...642A..12M}, Parker Solar Probe \citep[][]{2016SSRv..204...49B}, STEREO \citep{2008SSRv..136..487B}, and WIND \citep{1995SSRv...71..231B}.

\begin{figure}[ht]
    \centering
	\includegraphics[width=1.0\columnwidth]{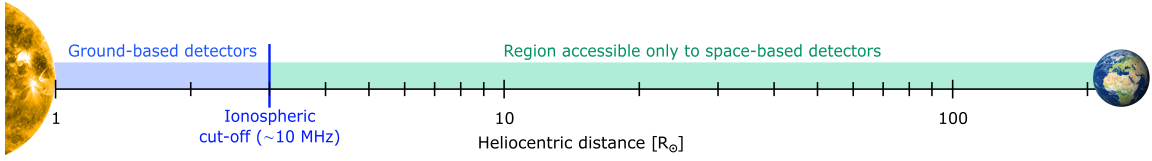}
    \caption{
    Schematic illustration of the distances between the Sun and the Earth covered by combining remote-sensing measurements from ground-based and space-based instruments, for the case of solar radio emissions (excited via the plasma emission mechanism). Space-based instruments are necessary to circumvent the ionospheric cut-off at $\sim$10~MHz (corresponding to approximately 3~$\mathrm{R_\odot}$, depending on the density profile), which prevents radio emissions of lower frequencies from reaching the ground. The SKA will be able to record the majority of the desired frequencies above the ionospheric cut-off.
    }
    \label{fig:cartoon}
\end{figure}

Synergies between the SKA and space-based detectors are particularly beneficial for studies of heliospheric turbulence. Radio-wave propagation effects impact lower-frequency radio burst emissions more than higher-frequency ones, so the diagnostic potential increases with decreasing frequency. Moreover, the emission frequency of solar radio bursts (emitted via the plasma emission mechanism) is directly related to the local heliospheric density, which decreases as a function of heliocentric distance, meaning that the frequency of solar radio bursts reflects the radio source distance from the Sun. Specifically, lower-frequency radio burst emissions are excited at larger heliocentric distances than their higher-frequency counterparts. Therefore, combining SKA observations, which will cover the distances closer to the Sun, with space-based observations at lower frequencies, which cover distances further out, will enable the diagnosis of heliospheric turbulence in the entire space between the Sun and Earth (and beyond). The heliocentric distances that will be covered when combining remote-sensing measurements from ground-based instruments (like the SKA) with those of space-based instruments are illustrated in Fig.~\ref{fig:cartoon}. For comparison, the plasma frequency near Earth (at $215$~$\mathrm{R_\odot}$) is on the order of a few kHz. Another benefit of synergies between the SKA and spacecraft is that, along with the turbulence properties inferred from their remote-sensing measurements, spacecraft can also make in-situ measurements of the heliospheric density turbulence that can complement and be compared with remote-sensing analyses, adding additional data points in the effort to characterise the turbulence across the heliosphere in a holistic manner \citep[e.g.,][]{2023ApJ...956..112K}.

\section{Summary} \label{sec:summary}
Turbulence, which leads to energy transfer and particle acceleration in astrophysical plasmas, is ubiquitous in the universe. The heliosphere is a natural laboratory, and one of the few astrophysical bodies for which we have statistically-enormous and diverse datasets, allowing us to probe turbulence using both remote-sensing and in-situ measurements, gaining understanding on how other active stars govern their stellar environment and astrospheres.

Radio photons are particularly affected by turbulence in the heliosphere, and are thus excellent diagnostic tools of its properties. In this chapter, we presented the variety of radio emissions that can be used to probe heliospheric turbulence, all of which the SKA will observe. These radio emissions, measured at a large range of frequencies, include those from extra-solar sources, solar sources (i.e., solar radio bursts), and synthetic sources. We also demonstrated that the combination of data from all the various radio sources enables the diagnosis of the heliospheric density inhomogeneities at any heliocentric distance, including those very close to the Sun that cannot be probed in situ by spacecraft. 

Due to its advanced observing abilities, and particularly its unprecedented sensitivity combined with high-resolution imaging spectroscopy observations, the SKA will be a prominent radio observatory for studies of heliospheric turbulence, as they require detailed observations of faint and fine radio structures. Furthermore, we highlighted the multi-faceted ways in which synergies with other radio detectors---both ground- and space-based---will complement SKA observations and enable the holistic characterisation of heliospheric turbulence.

\section{Acknowledgements}
N. Chrysaphi was supported by the Space It Up! project, funded by the Italian Space Agency (ASI) and the Ministry of University and Research (MUR), under contract No. 2024-5-E.0--CUP I53D24000060005.
E.P.K. acknowledges financial support from the STFC/UKRI (grant ST/Y001834/1) and the Leverhulme Trust (Research Fellowship RF-2025-357).

\bibliographystyle{abbrvnat-maxbibnames4}
\bibliography{chapter} 

\end{document}

%% file: journal-names.tex
\newcommand{\actaa}{Acta Astron.} 
\newcommand{\araa}{ARA\&A} 
\newcommand{\aar}{A\&ARv} 
\newcommand{\aapr}{A\&ARv} 
\newcommand{\ab}{Astrobiol.} 
\newcommand{\aj}{AJ} 
\newcommand{\apj}{ApJ} 
\newcommand{\apjl}{ApJL} 
\newcommand{\apjs}{ApJSS} 
\newcommand{\ao}{Appl. Opt.} 
\newcommand{\apss}{Astro. \& Space Sci.} 
\newcommand{\aap}{A\&A} 
\newcommand{\aaps}{A\&AS.} 
\newcommand{\baas}{Bull. Am. Astron. Soc.} 
\newcommand{\caa}{Chinese A\&A} 
\newcommand{\cjaa}{Chinese J. A\&A} 
\newcommand{\cqg}{Class. Quantum Gravity} 
\newcommand{\gal}{Galaxies} 
\newcommand{\gca}{Geo. Cosmo. Acta} 
\newcommand{\icarus}{Icarus} 
\newcommand{\jcap}{JCAP} 
\newcommand{\jgr}{J. Geophys. Res.} 
\newcommand{\jgrp}{J. Geophys. Res. Planets} 
\newcommand{\jqsrt}{J. Quant. Spectrosc. Radiat. Transf.} 
\newcommand{\memsai}{Mem. SAIt} 
\newcommand{\mnras}{MNRAS} 
\newcommand{\nat}{Nature} 
\newcommand{\nastro}{Nat. Astron.} 
\newcommand{\ncomms}{Nat. Commun.} 
\newcommand{\nphys}{Nat. Phys.} 
\newcommand{\na}{New Astron.} 
\newcommand{\nar}{New Astron. Rev.} 
\newcommand{\physrep}{Phys. Rep.} 
\newcommand{\pra}{Phys. Rev. A} 
\newcommand{\prb}{Phys. Rev. B} 
\newcommand{\prc}{Phys. Rev. C} 
\newcommand{\prd}{Phys. Rev. D} 
\newcommand{\pre}{Phys. Rev. E} 
\newcommand{\prx}{Phys. Rev. X} 
\newcommand{\prl}{Phys. Rev. Let.} 
\newcommand{\psj}{Planet. Sci. J.} 
\newcommand{\planss}{Planet. Space Sci.} 
\newcommand{\pnas}{Proc. Natl Acad. Sci. USA} 
\newcommand{\procspie}{Proc. SPIE} 
\newcommand{\pasa}{PASA} 
\newcommand{\pasj}{PASJ} 
\newcommand{\pasp}{PASP} 
\newcommand{\rmxaa}{RMXAA} 
\newcommand{\sci}{Science} 
\newcommand{\sciadv}{Sci. Adv.} 
\newcommand{\solphys}{Sol. Phys.} 
\newcommand{\sovast}{Soviet Ast.} 
\newcommand{\ssr}{Space Sci. Rev.} 
\newcommand{\uni}{Universe} 